\documentclass[aps,prb,superscriptaddress, twocolumn, showpacs]{revtex4-1}
\usepackage{graphicx}
\usepackage{dcolumn}
\usepackage{bm}
\usepackage{amsmath}
\usepackage[mathlines]{lineno}

\usepackage{color}
\usepackage{bm,graphicx,hyperref}
\hypersetup{%
  breaklinks = {true},
  citecolor = {blue},
  colorlinks = {true},
  linkcolor = {red},
}

\begin{document}

\title{Strain-induced gauge and Rashba fields in ferroelectric
  Rashba lead chalcogenide PbX (X=S, Se, Te) monolayers}

\author{Paul~Z.~Hanakata}
\affiliation{Department of Physics, Boston University, Boston, MA 02215}
\email{hanakata@bu.edu}

\author{A.~S.~Rodin} 
\affiliation{Centre for Advanced 2D
  Materials and Graphene Research Centre, National University of
  Singapore, 6 Science Drive 2, 117546, Singapore}

\author{Harold~S.~Park}
\affiliation{Department of Mechanical Engineering, Boston University, Boston, MA 
02215}

\author{David~K.~Campbell}
\affiliation{Department of Physics, Boston University, Boston, MA  02215}

\author{A.~H.~Castro Neto} 
\affiliation{Centre for Advanced 2D
  Materials and Graphene Research Centre, National University of
  Singapore, 6 Science Drive 2, 117546, Singapore}

\date{\today}
\begin{abstract}
  One of the exciting features of two-dimensional (2D) materials is
  their electronic and optical tunability through strain engineering.
  Previously we found a new class of 2D ferroelectric Rashba
  semiconductors PbX (X=S, Se, Te) with tunable spin-orbital
  properties.  In this work, based on our previous tight-binding (TB)
  results, we derive an effective low-energy Hamiltonian around the
  symmetry points that captures the effects of strain on the
  electronic properties of PbX. We find that strains induce gauge
  fields which shift the Rashba point and modify the Rashba
  parameter. This effect is equivalent to the application of in-plane
  magnetic fields. The out-of-plane strain, which is proportional to
  the electric polarization, is also shown to modify the Rashba
  parameter.  Overall, our theory connects strain and spin-splitting
  in ferroelectric Rashba materials, which will be important to
  understand the strain-induced variations in local Rashba parameters
  that will occur in practical applications.
\end{abstract}

\pacs{}

\maketitle
\section{Introduction} Monolayers and heterostructures of
two-dimensional (2D) materials with spin-orbit interaction offer
promise for observing many novel physical effects~\cite{geim2013van,
  novoselov-science-353-2016, manchon-NatMat-14-871-2015}. In
particular, it has been proposed that topological insulators or
semiconductors with Rashba interactions coupled with superconductors
may host Majorana fermions, which are potential building blocks for
topological quantum computers~\cite{sau-PRL-104-040502-2010,
  fu-PRL-100-096407-2008}.

In addition to 2D materials that exist in the hexagonal phase, such as
graphene and the transition metal dichalcogenides (TMDCs), 2D
materials with square lattices have been successfully
fabricated~\cite{chang-science-353-274-2016,
  moayed-NatComm-8-15721-2017}. Recently, the Rashba effect has been
observed in thin layers (6--20 nm) of lead sulfide
(PbS)~\cite{moayed-NatComm-8-15721-2017}, where an external electric
field is used to break the inversion symmetry. However, the
spin-splitting is not large. In our previous work based on density
functional theory (DFT) calculations, we found that lead chalcogenide
monolayers PbX (X=S, Se, Te) have large Rashba coupling
$\lambda\sim1$~eV\AA~in their non-centrosymmetric buckled
phase~\cite{hanakata-PRB-96-161401-2017}. In addition, the spin
texture can be switched in a non-volatile way by applying an electric
field or mechanical strain, which puts these materials into the family
of ferroelectric Rashba semiconductors
(FERSCs)~\cite{sante-AdvMat-25-1521-2013,
  sante-PRB-91-161401-2015}. This spin-switching mechanism has
recently been observed experimentally in thin films GeTe where the
surface is engineered to have either an inward or outward electric
polarization~\cite{rinaldi-NL-18-2751-2018}.

In reality, monolayers experience strains due to substrates, defects,
and so on, where local strains may change the electronic properties of
monolayers. Important examples of such effects are pseudo-Landau
levels in graphene blisters~\cite{levy-Science-329-544-2010} and band
gap shifts in biaxially strained
MoS$_2$~\cite{castellanos-NL-13-5361-2013}. Recently, spatial
variations of Rashba coupling due to variations in local electrostatic
potentials were reported in InSb~\cite{bindel-NatPhys-12-920-2016}. To
date, most theoretical studies of lead chalcogenide monolayers have
been based solely on DFT calculations~\cite{liu-NanoLett-15-2657-2015,
  wan-AdvMat-29-1521-2017}.  However, because DFT is limited to the
simulation of small systems, typically several nanometers, it is
difficult to model inhomogeneous strains over large spatial areas
using DFT.

In this paper, based on our previous tight-binding (TB)
model~\cite{hanakata-PRB-96-161401-2017, rodin-PRB-96-115450-2017}, we
develop a continuum model to predict strain-induced changes in the
spin and electronic properties of buckled PbX monolayers. We have also
performed DFT calculations to validate our TB predictions. Due to the
buckled structure of PbX, the angular dependence becomes important as
the relative angle between hybrid orbitals of the top and bottom layer
can change substantially~\cite{hanakata-PRB-96-161401-2017}. We note
that some studies on (non-buckled) SnTe and PbX (X=S, Se, Te)
rock-salt type materials have incorporated strain effects in the TB,
but did not include the changes in hopping parameters due to angle
changes~\cite{tang-NatPhys-10-964-2014, barone-physica-7-1102-2013}.
In contrast, our TB formulation incorporates the effects due to
changes in (i) bond distance and (ii) angle between nearest neighbors
as well as (iii) lattice vector deformation.

In the low-energy Hamiltonian, the biaxial (or uniaxial) strains can
be described as gauge fields, which are equivalent to, by minimal
coupling, the application of in-plane magnetic fields. The
out-of-plane strain is directly related to the out-of-plane
polarization and this also modifies the Rashba parameter.  Within this
framework we are able to quantify the Rashba fields in terms of the
strain fields.

\begin{figure*}
\includegraphics[width=16cm]{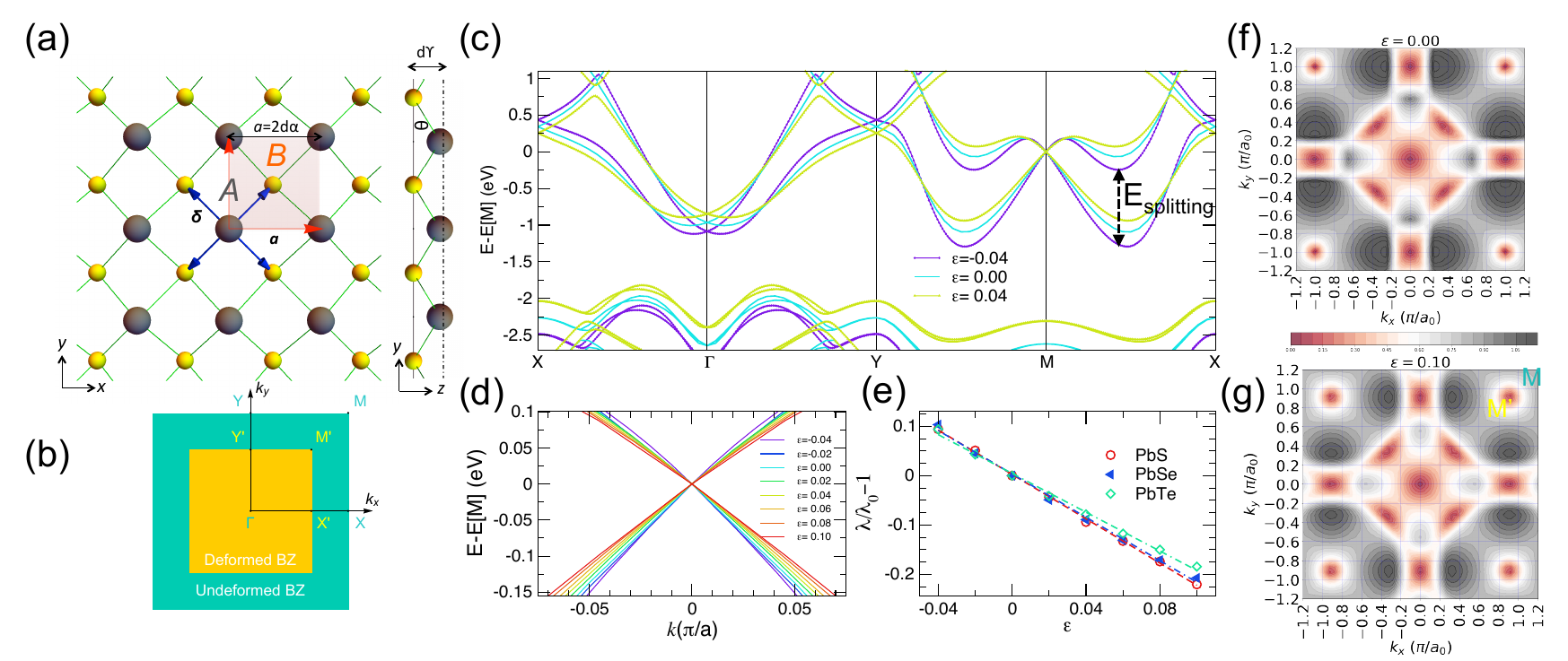}
\caption{(a) Schematic top and side views of a buckled $AB$
  monolayer. (b) Undeformed and deformed Brillouin zone as the monolayer
  is stretched in the $x$ and $y$ direction. (c) Representative band
  structures of strained PbS along symmetry points
  $X$-$\Gamma$-$Y$-$M$-$X$ and (d) close to $M$. (e) Relative change
  in the Rashba parameters obtained from DFT calculations as a function of
  strain $\epsilon$ for PbS, PbSe, and PbTe. Energy spin-splitting of
  PbS for isotropic strains of (f) $\epsilon=0.00$ and (g)
  $\epsilon=0.10$. It can be seen that the $M$ points are originally
  located at $|k_{x,y}|=\pi/a_0$ and shifted closer to the center
  under a strain of $\epsilon=0.10$.}
\label{fig:fig1}
\end{figure*}

\section{Tight-binding} Lead chalcogenide PbX (X=S, Se, Te) consists of
two atoms per unit cell, denoted by $A$ and $B$ atoms,
respectively. Lead is a heavy atom (Z(Pb)=82), and it is crucial for
creating large spin-orbit interaction (SOI). The schematic top and
side views of a buckled $AB$ lattice are shown in
fig.~\ref{fig:fig1}(a). $\pmb{a}$ is the unit lattice vector and
$\pmb{\delta}_j$ is the vector connecting atom $i$ and its $j$
neighbor. We denote the relaxed bond length between the neighboring
$A$ and $B$ atoms by $d$, the vector connecting $A$ and $B$ atoms in
the $(0, 0)$ unit cell $\pmb{\delta}_1=d (\alpha,\alpha,-\gamma)$
where $\alpha=\frac{\cos\theta}{\sqrt 2}$, $\gamma=\sin\theta$, and
$\theta$ is the buckling angle (with $\theta=0$ corresponding to a
flat lattice).

The bands near the Fermi level are mostly composed of $s$ and $p$
orbitals from both $A$ and $B$
atoms~\cite{hanakata-PRB-96-161401-2017}. The bands near the symmetry
points can be described within the TB framework including first
nearest neighbors and SOI. The full derivation of the TB model can be
found in our previous works~\cite{hanakata-PRB-96-161401-2017,
  rodin-PRB-96-115450-2017}, and thus we will only outline the
important parts; a more detailed derivation can be found in
Appendix~\ref{sec:tb}.

For the two atom $AB$ unit cell shown in Fig.~\ref{fig:fig1}(a), 
the relevant orbital basis involves
$\{s^A, p_x^A, p_y^A, p_z^A, s^B, p_x^B, p_y^B, p_z^B\}$.  
To write down the hopping matrix, we use the Slater-Koster
matrix elements for the orbitals of neighboring
atoms~\cite{slater-PR-94-1498-1954}. As we include the SOI,
$H_\mathrm{SOI} = T_\mathcal{X}\left(\frac{L_+\otimes s_-+L_-\otimes
    s_+}{2}+L_z\otimes s_z\right)$
(where $\mathcal{X}=A, B$), we will write our Hamiltonian in angular momentum
basis. The dimension of the total Hilbert space is $16\times16$ with
new basis of
$|\mu\rangle\rightarrow|m\rangle|m_{\rm orb}\rangle|s\rangle$, where
$m=\{|A\rangle, |B\rangle\}$ is the sublattice degree of freedom,
$m_{\rm orb}=\{|0, 0\rangle, |1, 1\rangle, |1, -1\rangle, |1,
0\rangle\}$
is the orbital angular momentum degree of freedom, and $s=\{(|+\rangle, |-\rangle\}$ is
the spin degree of freedom.

We found a Rashba-like dispersion near the $\Gamma$ and $M$ points
when the two sublattices are not
equivalent~\cite{hanakata-PRB-96-161401-2017,
  rodin-PRB-96-115450-2017}. In this paper, we develop a continuum
strain model describing changes in the Rashba dispersion near the $M$
point, and thus the Hamiltonian is expanded around the $M$ point
${\bf k}=(\pi/a, \pi/a)$. Exactly at $M$ [$q=0$], the Hamiltonian
decomposes into several uncoupled blocks and the wave function of the
conduction band is given by
$|\Psi^\pm\rangle_{mn}
=c_0|m\rangle\otimes|1,\pm1\rangle\otimes|\mp\rangle+c_1|m\rangle\otimes|1,0\rangle\otimes|\pm\rangle\pm
ic_2|n\rangle\otimes|1,\mp1\rangle\otimes|\mp\rangle$,
with $c_0$, $c_1$, and $c_2$ being real
numbers~\cite{hanakata-PRB-96-161401-2017,
  rodin-PRB-96-115450-2017}. The Hamiltonian for the valance band can
be obtained by interchanging $m$ and $n$.

Projecting the Hamiltonian onto the conduction band subspace we obtain the effective Rashba-like Hamiltonian 
\small
\begin{align}
	H_\mathrm{eff}^{mn} &= \lambda\left[\left(\mathbf{q}\times\pmb{\sigma}\right)\cdot\hat z\right]:\begin{pmatrix}
		|\Psi^+\rangle_{mn}
		\\
		|\Psi^-\rangle_{mn}
	\end{pmatrix}\,,
	\label{eq:H_eff_M}
\end{align}
\normalsize where $\mathbf{q}$ is the momenta,
$\pmb{\sigma}=(\sigma_x, \sigma_y, \sigma_z)$,
$\lambda\equiv a\sin2\theta \Delta c_1c_2$ is the Rashba parameter,
and $\Delta=V_{pp\sigma}-V_{pp\pi}$.  The coefficients $c_0, c_1, c_2$
can be obtained from the DFT calculations. Since we know the buckling
angle $\theta$ we can can evaluate $\Delta$. All of the relevant
(unstrained) parameters are tabulated in Appendix~\ref{sec:dft}.

\section{Strain-induced gauge fields} 
\begin{figure*}
\includegraphics[width=12cm]{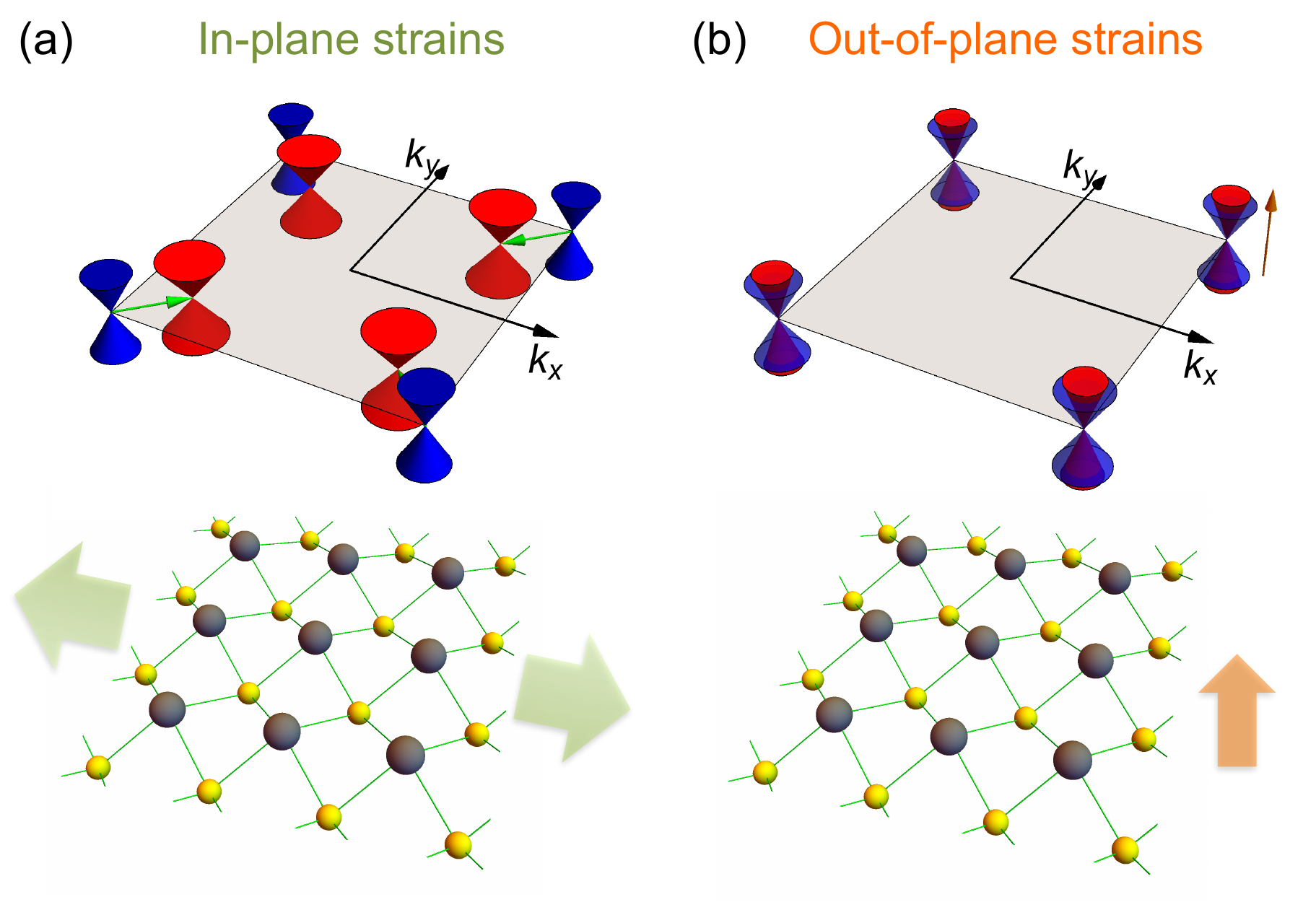}
\caption{Schematic changes in the Rashba dispersions due to (a)
  in-plane strains and (b) out-of-plane strains. The linear Rashba
  dispersions at the $M$ for unstrained systems are colored blue.
  Under positive in-plane strains, the Rashba points shift closer to
  $\Gamma$ and the strength of Rashba parameters decrease (smaller
  slope) with increasing strains. On the other hand, under
  out-of-plane strain, the strength of Rashba parameters increases
  with increasing uniaxial out-of-plane strain while the Rashba points
  do not shift. }
\label{fig:fig2}
\end{figure*}

Since the SOI is independent of lattice distortions, in this
derivation we will focus on the spinless Hamiltonian and then
reintroduce the spin terms. We will focus on the conduction band only,
as the changes in valence band should be similar.

Under deformation a vector connecting two points in a unit cell $i$
can be approximated as
$\pmb{r}'_j-\pmb{r}'_i\simeq \pmb{\delta}_j +\pmb{\delta}_j\cdot\nabla
\pmb{u}(\pmb{r}_i)$,
where $\pmb{u}=(u_x, u_y, u_z)$ is the displacement vector, and
$\nabla \pmb{u}=\tilde{\pmb{\epsilon}} + \tilde{\pmb{\omega}}$. In
this work we focus on deformation that does not involve local rotation
$\tilde{\pmb{\omega}}=0$.  Similarly, between two lattice vectors
$\pmb{R}'_j-\pmb{R}'_i\simeq \pmb{a}^i +\pmb {a}^i\cdot\nabla
\pmb{u}(\pmb{R}_i)$.

Alterations in bond distance will result in changes in the hopping
energies. Since studies of lead chalcogenides under strain are very
limited, we follow the Wills-Harrison's
argument~\cite{harrison2004elementary} and assume that the hopping
energy $t\propto r^{-\beta_{\mu\nu}}$. Similar considerations also
have been used for strained TMDCs~\cite{cazalilla-PRL-113-077201-2014,
  rostami-PRB-92-195402-2015, fang2017electronic} and
phosphorene~\cite{jiang-PRB-91-235118-2015,
  sisakht-PRB-94-085417-2016}. Note that the hopping matrix derived
from Slater-Koster has angular dependence and these relative angles
should change due to strain. Assuming the hopping matrix depends on
bond distance only, the modified hopping parameter, in terms of the
strain tensor $\tilde{\pmb{\epsilon}}$, is
$t'_{ij, \mu\nu}(\delta_{ij}) \simeq t_{ij, \mu\nu}(1- \beta_{\mu\nu}
\frac{1}{d^2}\pmb{\delta}_j\cdot\tilde{\pmb{\epsilon}}\cdot{\pmb{\delta}}_j)$~\cite{cazalilla-PRL-113-077201-2014,
  rostami-PRB-92-195402-2015}.  This approximation
is also the case for graphene, where the hopping modulation is
approximated as
$t'(\pmb{\delta}_{ij}) = te^{-\beta (|\pmb{\delta}_{ij}|/d-1)}$.  In
particular, this approximation works well for {\it flat} graphene
under strain because the angle between $p_z$ orbitals does not
change. The angular dependence becomes more important when
deformations, such as nanobubbles and kirigami patterns, create large
curvature (bending)~\cite{qi-PRB-90-125419-2014,
  qi-PRB-90-245437-2014}. In buckled lead chalcogenides, however, the
relevant hopping terms for the Rashba dispersion depend on the
buckling angle even in the simple case of biaxial
strains~\cite{hanakata-PRB-96-161401-2017}. Thus we will include this
angular dependence, and we will show that this is important to capture
the changes in Rashba coupling with uniaxial strain.

Let the unstrained vector connecting an atom $A$ and its neighbor be
defined as $\pmb{\delta}_j=(x, y, z)$ and the equilibrium distance
$r=d$. Here we show the derivation for $t_{p_xp_z}$, while the others
can be found by following the same procedure.  We assume
$\Delta(r')=\Delta_0\left(\frac{r}{r'}\right)^\beta$ and we expect
$\beta\approx 3$~\cite{harrison2004elementary}. In Cartesian
coordinates the strained hopping is given by
$t_{p_xp_z}(x',y',z')=\frac{x'z'}{r'^2}\Delta_0\left(\frac{r}{r'}\right)^\beta$,
and by Taylor expansion we obtain, \small
\begin{widetext}
\begin{equation}
\delta t_{ij, p_xp_z}(x', y', z') \simeq-t_{ij, p_xp_z}(x, y, z) \Big(\Big[(2+\beta)-(r/x)^2\Big]\frac{1}{r^2}{\bf x}\cdot({\bf x}'-{\bf x})-\Big[2+\beta\Big]\frac{1}{r^2}{\bf y}\cdot({\bf y}'-{\bf y})-\Big[(2+\beta)-(r/z)^2\Big]\frac{1}{r^2}{\bf z}\cdot({\bf z}'-{\bf z})\Big). 
\end{equation}
\end{widetext}
\normalsize Within the strain approximation
${\bf x}'-{\bf
  x}=\hat{x}\cdot\tilde{\pmb{\epsilon}}\cdot{\pmb{\delta}}_j$.
If we alter only the bond distance while keeping the angle constant,
we will get the same expression as above when angular effects are
assumed to be negligible.

The interlattice-spinless Hamiltonian in reciprocal space can be written as 
\small
\begin{widetext}
\begin{align}
H^{\rm int}_{\rm orb}({\bf k})=&\sum_{\mu, \nu}\sum_{\langle ij \rangle}(t_{ij, \mu\nu} +\delta t_{ij, \mu\nu})e^{i{\bf k}\cdot\pmb{\Delta}_j(1+\tilde{\pmb{\epsilon}})}c^{\dagger}_{i, {\bf k}, \mu}c_{j, {\bf k}, \nu}+h.c.\nonumber\\
                =&\underbrace{\sum_{\mu, \nu}\sum_{\langle i, j\rangle}t_{ij, \mu\nu}e^{i{\bf k}\cdot\pmb{\Delta}_j}c^{\dagger}_{i, {\bf k}, \mu}c_{j,  {\bf k}, \nu}}_{H_0}+\underbrace{\sum_{\mu, \nu}\sum_{\langle i, j\rangle} it_{ij, \mu\nu}{\bf k}\cdot\tilde{\pmb{\epsilon}}\cdot\pmb{\Delta}_j e^{i{\bf k}\cdot\pmb{\Delta}_j}c^{\dagger}_{i, {\bf k}, \mu}c_{j, {\bf k}, \nu}}_{H^{(1)}}+\underbrace{\sum_{\mu, \nu}\sum_{\langle i, j\rangle}\delta t_{ij, \mu\nu}e^{i{\bf k}\cdot\pmb{\Delta}_j}c^{\dagger}_{i, {\bf k},\mu}c_{j, {\bf k},\nu}}_{H^{(2)}}+\mathcal{O}(\epsilon^2), 
\end{align}
\end{widetext}
\normalsize where $\langle ij\rangle$ is the sum over nearest neighbor
pairs and $\pmb{\Delta}_j={\bf R}_j-{\bf R}_i$. The first term $H_0$ is
the unstrained Hamiltonian, $H^{(1)}$ is the correction due to lattice
deformation, and $H^{(2)}$ is the correction from the altered hopping
parameter due to changes in both the interatomic distance and angle between
orbitals. 

\section{Homogenous isotropic strains} We start with a simple deformation with no shear \small
$\tilde{\pmb{\epsilon}}=\begin{pmatrix}
\epsilon_{xx} & 0 & 0  \\
0 & \epsilon_{yy} & 0\\
0 & 0 & \epsilon_{zz}
\end{pmatrix}.$
We will focus on the matrix elements that are relevant to the
conductions band, such as $|A\rangle|1, 0\rangle$ and
$|B\rangle|1,1\rangle$. In the {\it angular momentum} basis, the
correction from $H^{(1)}$ and $H^{(2)}$ at $M$ is given by \small
\begin{widetext}
\begin{align}
_A\langle 1, 0|H^{(1)}|1, 1\rangle_B=&a_0\sqrt{2}\alpha_0\Delta_0\gamma_0\Big[\epsilon_{xx}\pi/a_0+q_x\epsilon_{xx}-i\epsilon_{yy}\pi/a_0-iq_y\epsilon_{yy}\Big]\nonumber\\
_A\langle 1, 0|H^{(2)}|1, 1\rangle_B=& -a_0\sqrt{2}  \alpha_0 \gamma_0 \Delta_0 \alpha_0^2(2+\beta)\Big[ (\epsilon_{xx} + f_1 \epsilon_{yy} + f_2\epsilon_{zz})q_x- (f_1\epsilon_{xx} + \epsilon_{yy} + f_2 \epsilon_{zz})iq_y\Big]\,
\label{eq:delta_H}
\end{align}
\end{widetext}
\normalsize where
$\epsilon_{ij}=\frac{1}{2}\Big(\frac{\partial u_i}{\partial
  x_j}+\frac{\partial u_j}{\partial x_i} +\frac{\partial u_l}{\partial
  x_i} \frac{\partial u_l}{\partial x_j}\Big)$,
$f_1=1-\frac{1}{\alpha_0^2(2+\beta)}$ and
$f_2=\frac{\gamma_0^2}{\alpha_0^2}-\frac{1}{\alpha_0^2(2+\beta)}$. Note
that $a_0, \alpha_0, \beta_0, \gamma_0, \Delta_0$ are the {\it
  unstrained} geometrical and hopping parameters.  $H^{(1)}$ is
independent of the $z$ direction strains (e.g $\epsilon_{xz}$) because
the lattice vector ${\bf R}$ and ${\bf k}$ are two-dimensional.
Because of the symmetry of $M$, we found that the first correction at
$M$ due to bond alterations is first order in $\epsilon$ {\it and}
momentum $q$. In graphene, the first correction from hopping
modulation that is linear in $\epsilon$ (but not proportional to $q$)
is not zero~\cite{guinea-NatPhys-6-30-2010, juan-PRB-88-155405-2013,
  masir-ssc-175-76-2013}. We have to include the contributions of
$H^{(1)}$ up to first order in $q$ as well because in $H^{(2)}$
($\beta$-dependent term) we keep terms up to first order in $q$ and
$\epsilon$.

To obtain $\beta$ we will consider an isotropic strain
$\epsilon\cdot 1_{3\times3}$. Notice that the change in low-energy
Hamiltonian of Eq.~\ref{eq:H_eff_M} due to $H^{(1)}$ and $H^{(2)}$ at
$M$ can be written as gauge potentials, \small
\begin{align}
H_\mathrm{eff} &=-i\lambda_0\begin{pmatrix}
0&(q_x-iq_y) + \pmb{A}_1 + \pmb{A}_2\\
(q_x+iq_y) +  \pmb{A}_1^* + \pmb{A}_2^* &0
\end{pmatrix}\,.
\label{eq:Heffmod}
\end{align}
\normalsize
where 
\small
$\pmb{A}_1= \begin{pmatrix} \epsilon\pi/a_0  + \epsilon\,q_x\\
-i\epsilon\pi/a_0 -i\epsilon\,q_y
\end{pmatrix}\quad {\rm and}\quad$
$\pmb{A}_2 = -\beta\begin{pmatrix}\epsilon\,q_x\\
-i\epsilon\,q_y
\end{pmatrix}\,$
\normalsize where we have used $2\alpha_0^2+\gamma_0^2=1$ to simplify
$\pmb{A}_1, \pmb{A}_2$ and $\lambda_0$ is the unstrained Rashba
parameter.

$\pmb{A}_2$ and the second term of $\pmb{A}_1$ are {\it proportional}
to $q$. This modifies the strength of Rashba parameter
$\frac{\lambda}{\lambda_{0}}-1\simeq(1-\beta)\epsilon$. This
alteration in the Rashba term is similar to the modification of Fermi
velocity in graphene~\cite{juan-PRL-108-227205-2012,
  juan-PRB-88-155405-2013, masir-ssc-175-76-2013}. 

We next present our DFT results to validate our TB
predictions. Details of DFT calculations and the unstrained
geometrical parameters of buckled PbS, PbSe, and PbTe can be found in
Appendix~\ref{sec:dft}. Strains are applied to the relaxed buckled phase. In order to
find the effects that come from changes in bond distance only, we
deformed the monolayer in the DFT simulations by changing the bond
distance while keeping the angle constant. The lattice vectors and
atomic positions are not relaxed under this deformation. The Rashba
parameters $\lambda$ are obtained by taking the derivative of the
energy dispersion in the vicinity of the $M$ point, $|q|<0.1\pi/a$. Under
isotropic deformations, we found that $\lambda$ at $M$ decreases with
increasing strain (weakening of the hopping interaction), as expected
from Eq.~\ref{eq:Heffmod}, shown in fig.~\ref{fig:fig1}(c)-(e). A direct
comparison between DFT results and TB with strain-included allows us
to extract $\beta$. By fitting DFT data points to a straight line, we
obtained $\beta=3.25, 3.20, 2.97$ for PbS, PbSe, and PbTe,
respectively (fig.~\ref{fig:fig1}(e)). We see that the value of
$\beta$ would be different if the lattice deformation correction was
not included.

As we stretch the lattice, the Brillouin zone (BZ) will shrink, and the corner of the
BZ ($M$ point) will shift as
$(\frac{\pi}{a_0},
\frac{\pi}{a_0})\rightarrow(\frac{\pi}{a_0(1+\epsilon)},
\frac{\pi}{a_0(1+\epsilon)})\simeq(\frac{\pi}{a_0}(1-\epsilon),
\frac{\pi}{a_0}(1-\epsilon))$,
where $a_0$ is the undeformed lattice constant.  For positive strains,
the $M$ point shifts towards the $\Gamma$ point (relative to the
undeformed BZ), shown in fig.~\ref{fig:fig1}(b). In our modified TB
model, the $M$ point is displaced due to the first term of the lattice
deformation correction $\pmb{A}_1$ (see Eq.~\ref{eq:Heffmod}). The
momentum shifts due to lattice deformations are also found in
graphene~\cite{kitt-PRB-85-115432-2012}. The changes in Rashba
dispersion and its locations due to strains are illustrated in
fig.~\ref{fig:fig2}.

To show the momentum shifts relative to the undeformed (reference)
state, we plot the energy spin-splitting at the conduction band of PbS
obtained from the DFT results as a function of $k_x, k_y$, shown in
fig.~\ref{fig:fig1}(f) and (g). Note that momenta are in units of
$\pi/a_0$. Originally the $M$ points are located at
$|k_{x, y}|=\pi/a_0$ and are shifted closer to $\Gamma$
($|k'_{x, y}|\approx0.9\pi/a_0$) when an isotropic strain of
$\epsilon=0.10$ is applied. The momentum shift is linear with strains
${\bf k}\cdot\tilde{\pmb{\epsilon}}$, consistent with several previous
works~\cite{kitt-PRB-85-115432-2012, masir-ssc-175-76-2013}. This
Rashba-point shift due to strains is equivalent to applying
in-plane-magnetic fields $\mathbf{B}_{\rm ex}$ to the system,
\begin{equation}
	H =\lambda_0\left[\left(\mathbf{q}-\frac{e\mathbf{A}_{\rm ex}}{c}\right)\times\pmb{\sigma}\right]\cdot \hat{z}+m_\perp\sigma_zB_\perp+m_\parallel\mathbf{B}_\parallel\cdot\sigma_\parallel\,
\label{eq:magnetic}
\end{equation}
where $m_\perp=-\mu_B(c_1^2-2c_2^2)$,
$m_\parallel=-\mu_Bc_1(\frac{c_0}{\sqrt{2}}+c_1+c_0)$, and $\mu_B$ is
the Bohr magneton.  For completeness the derivation of
Eq.~\ref{eq:magnetic} is included in 
Appendix~\ref{sec:MagneticField}. As an illustration, we can choose an
external field of $\mathbf{A}_{\rm ex}=(0, 0, B_xy-B_yx)$, upon which
the in-plane magnetic field is given by
$\mathbf{B}_{\rm ex}=\nabla\times \mathbf{A}_{\rm ex}=(B_x, B_y, 0)$.
Since the Bohr magneton is small, in order to get a similar effect of
2\% strain using magnetic fields, one has to apply external magnetic
fields with an approximate strength of
$|B_{\rm
  ex}|\sim\frac{\sqrt{2}\,0.02\,\pi\lambda_0}{a_0\,m_\parallel}\approx
600$ Tesla (by Eq.~\ref{eq:Heffmod} and Eq~\ref{eq:magnetic}).

\begin{figure}
\includegraphics[width=8cm]{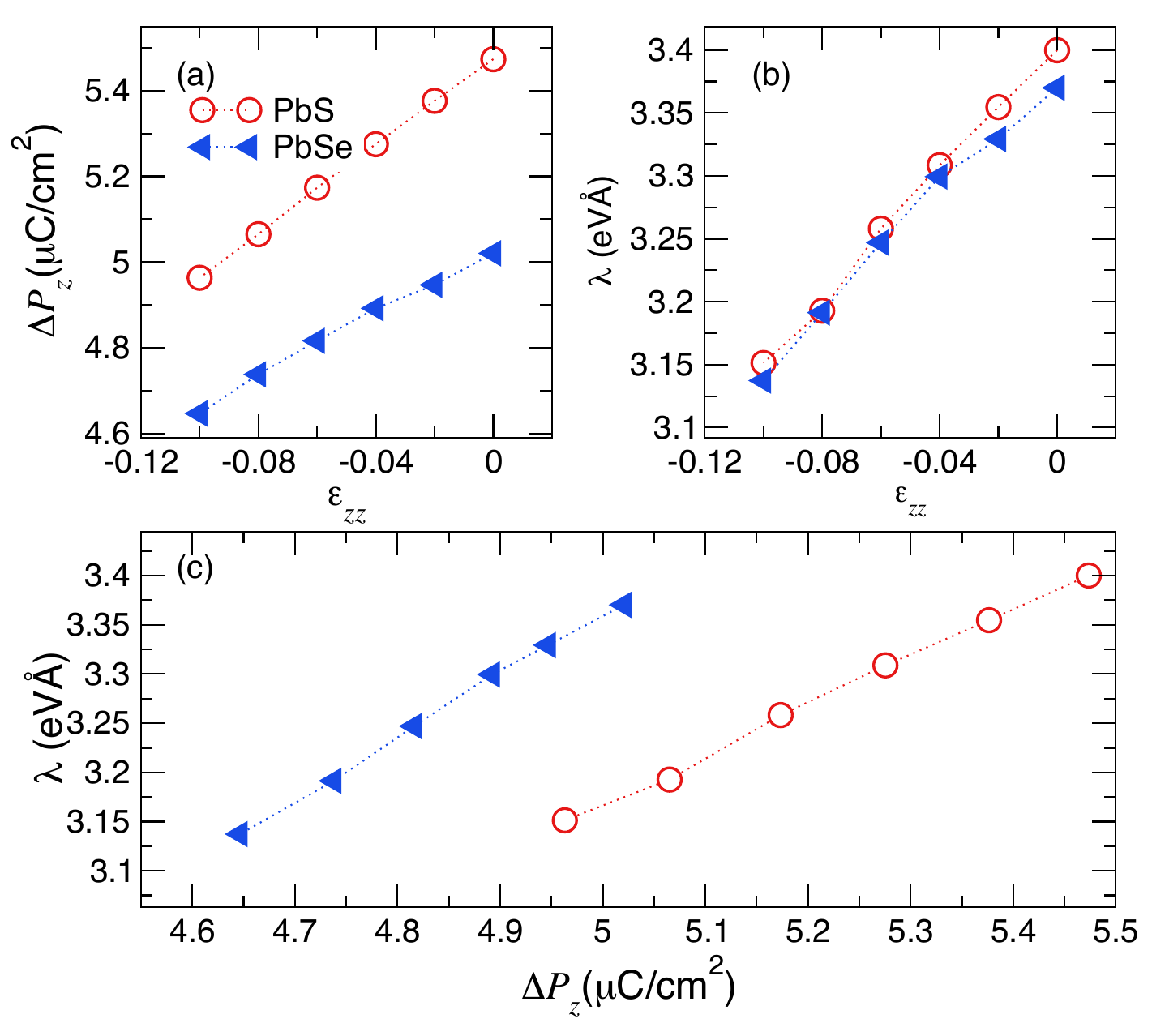}
\caption{(a) Out-of-plane polarization $\Delta\vec{\mathcal{P}}_z$ as
  a function of out-of-plane strain $\epsilon_{zz}$. (b) Linear
  relationship between $\lambda$ and $\epsilon_{zz}$ which is
  consistent with TB predictions. (c) Rashba parameter $\lambda$ as a
  function of $\Delta\vec{\mathcal{P}}_z$. All data points are obtained
  from the DFT calculations.}
\label{fig:fig3}
\end{figure}

\section{Electric polarization and Rashba field}
Proposals have been made to change the spin texture (i.e. sign of
$\lambda$) by changing the electric
polarization~\cite{sante-AdvMat-25-1521-2013,
  liebmann-AdvMat-28-560-2016, leppert-jpcl-7-3683-2016,
  qihang-NL-13-5264-2013}. Rinaldi {\it et al.} found that the
spin-texture in FERSC GeTe films indeed depends on the locations of
the atoms on the surface, which dictate the direction of the electric
polarization~\cite{rinaldi-NL-18-2751-2018}. In DFT simulations of
SnTe thin films, which have a structure similar to PbX, it also has
been shown that near the vacuum one of the atomic species buckles
outward while the other species buckles
inward~\cite{qian-NR-8-967-2015}. While the proportionality between
Rashba parameter and spontaneous electric polarization is well known,
it will be useful to understand this mechanism in PbX from a
microscopic view, where the changes in Rashba parameters can be
understood in terms of interactions between atoms and the external
applied strains. We will show that our strain-dependent TB model
captures how the out-of-plane strain, which is proportional to the
out-of-plane polarization, modifies the Rashba fields.

By the modern theory of polarization, the electric polarization is
given by\cite{vanderbilt-PRB-47-1651-1993}
$\vec{\mathcal{P}}=\frac{1}{V}\sum_{\tau}q^{\rm{ion}}_{\tau}{\bf
  R}_{\tau}-\frac{2i\rm{e}}{(2\pi)^3}\sum_{n}^{\rm{occ}}\int_{BZ}d^3{\bf
  k}e^{-i\vec{k}\cdot{\bf R}}\Big\langle \Psi_{n{\bf
    k}}\Big|\frac{\partial \Psi_{n{\bf k}}}{\partial {\bf
    k}}\Big\rangle$,
where $q_\tau$ is the ionic charge plus the core electrons,
${\bf R}_{\tau}$ is the position of ions, $V$ is the unit cell volume,
$\rm{e}$ is the elementary charge, $n$ is the valence band index,
${\bf k}$ is the wave vector, and $\Psi_{n{\bf k}}$ is the electronic
wave function. The first term is the contribution from core electrons
and ions, and the second term is the electronic contribution defined
as the adiabatic flow of current, which can be calculated from the
Berry phase (BP)~\cite{vanderbilt-PRB-47-1651-1993}. The spontaneous
polarization is calculated by taking the difference between the
polarization of the polar (buckled) state and the non-polar
(reference) state,
$\Delta \vec{\mathcal{P}}=\vec{\mathcal{P}}_{\rm
  polar}-\vec{\mathcal{P}}_{\rm non-polar}$.
We estimate the thickness to be 0.5 nm in order to compare the
polarizations to typical bulk ferroelectrics. Details can be found in
Appendix~\ref{sec:ferro}.  In the DFT simulations we distort the ions
in the $z$ direction while keeping the in-plane lattice vectors fixed
at the relaxed buckled values. We report only the spontaneous
polarizations of PbS and PbSe, as PbTe is
metallic~\cite{hanakata-PRB-96-161401-2017}. A modified Berry phase
calculation is needed to evaluate polarization of ferroelectric
metals~\cite{filippetti-NatCom-7-11211-2016}; however this is beyond
the scope of our present study.

From the DFT results we found that the core electronic plus ionic and
the electronic contribution (BP) are proportional to the distance
between Pb and X (X=S, Se) in the $z$ direction (plotted in
Appendix~\ref{sec:ferro}). This gives a proportionality between
$\Delta \vec{\mathcal{P}}_z$ and $\epsilon_{zz}$, as shown in
fig.~\ref{fig:fig3}(a). Compressing the monolayer in the $\hat{z}$
with strain $\epsilon_{zz}<0$ results in a decrease in $\lambda$,
shown in fig.~\ref{fig:fig3}(b). This is {\it opposite} to the case of
isotropic deformation (see fig.~\ref{fig:fig1}(e)). This result is
consistent with TB predictions. In the previous discussion, we found
that increasing bond distance ($\epsilon>0)$ generally weakens the
hopping interaction and thus decreases $\lambda$. Using relaxed
geometrical parameters (i.e buckling angle $\theta_0$) and from
Eq.~\ref{eq:delta_H}, $\lambda$ is expected to decrease with
compressive strain in the $\hat{z}$ as $f_2$ is negative. We also want
to note that there is no gauge-field $\pmb{A}_1$ since ${\bf k}$ is
two-dimensional, and thus $M$ is not shifted. The changes in Rashba
dispersion and its locations due to out-of-plane strains are
illustrated in fig.~\ref{fig:fig2}(b). Notice that not including the
angular dependence in the hopping correction will not capture this
effect. The inclusion of the angular dependence is particularly
important for the PbX monolayer due to its buckled nature. Overall,
this suggests that the out-of-plane internal electric polarization
acts as an in-plane gauge field in the low-energy
Hamiltonian. Assuming {\it small} strains, we found that
$\lambda\propto |\vec{\mathcal{P}}_z|$. This result is important as it
establishes a direct relationship between the Rashba field and the
out-of-plane polarization which is also proportional to the
out-of-plane strain $\epsilon_{zz}$. Recently, several works have also
studied strain-induced piezoelectricity in boron
nitride~\cite{droth-PRB-94-075404-2016} and
TMDCs~\cite{rostami-npj2d-2-15-2018}. Several experimental works use
out-of-plane magnetic fields (parallel to the polar axis of Rashba
materials) to measure the Rashba parameter as the Landau level
spectrum changes with the strength of the Rashba
parameter~\cite{bindel-NatPhys-12-920-2016,
  bordacs-PRL-111-166403-2013}. One could also use this experimental
approach to detect variations in the Rashba parameter in PbX due to
out-of-plane strains.

\section{Conclusions} 
We have developed a TB model where the electronic changes in PbX can
be described within continuum mechanics. We found the scaling exponent
that modifies the hopping parameter to be $\beta\simeq3$. In the
low-energy Hamiltonian, the effect of strains can be described as
gauge fields, which are equivalent to, by minimal coupling,
application of an in-plane magnetic field. Our theory describes how
the location of the Rashba point and the strength of the Rashba field
can be engineered by applying strains. The out-of-plane strain in
particular is directly related to the out-of-plane
polarization. Within this framework we are able to understand the
connection between the Rashba and ferroelectricity.

Our strain-dependent TB model should be applicable for calculating the
effects of inhomogeneous strain on the spatially-resolved Rashba
fields over a large region, whereas this calculation would not be
feasible within a reasonable time using a DFT approach. Employing
classical atomistic simulations (e.g. molecular dynamics) together
with strain-dependent TB will be an efficient tool for studying larger
and more realistic systems with strain modulation due to substrates,
indentors~\cite{levy-Science-329-544-2010,
  castellanos-NL-13-5361-2013, georgi-NL-17-2240-2017} or geometrical
cuts~\cite{qi-PRB-90-245437-2014, hanakata-Nanoscale-8-458-2016}.This
will open possibilities of using lead chalcogenides for strain and
electric-controlled spintronic devices.
\bibliography{biball}
\begin{acknowledgments}
P.Z.H. developed the theory, wrote the paper, and performed
the DFT calculations. A.S.R. contributed to the analytical
work. H.S.P., D.K.C., and A.C.H.N. supervised the research.
We thank Vitor M. Pereira for useful comments and discussions.
P.Z.H., H.S.P., and D.K.C. acknowledge support
by the Boston University Materials Science and Engineering
Innovation Grants. P.Z.H, H.S.P., and D.K.C are grateful for
computing resources provided by theBostonUniversity Shared
Computing Cluster.
\end{acknowledgments}

\appendix
\section{Methods}
\subsection{Computational Details}
\label{sec:dft}
To validate our tight-binding predictions we performed density
functional theory (DFT) calculations implemented in the {\sc Quantum
  ESPRESSO} package~\cite{QE-2009}.  We employed projector
augmented-wave (PAW) type pseudopotentials with Perdew-Burke-Ernzerhof
(PBE) within the generalized gradient approximation (GGA) for the
exchange and correlation functional
with~\cite{pbe-PRL-77-3865-1996}. The Kohn-Sham orbitals were expanded
in a plane-wave basis with a cutoff energy of 100 Ry and a charge
density cutoff of 200 Ry. The cutoff was chosen following the
suggested minimum cutoff in the pseudopotental file.  A $k$-point grid
sampling was generated using the Monkhorst-Pack scheme with
16$\times$16$\times$1 points~\cite{monkhorstPRB1976}. A vacuum of
20~\AA~was used. The relaxed structures of PbS, PbSe, and PbTe were
obtained by relaxing the ionic positions and the lattice vectors.  A
convergence threshold on total energy of $10^{-5}$ eV and a
convergence threshold on forces of 0.005 eV/${\rm \AA}^{-1}$ were
chosen. Lattice vectors are relaxed until the stress is less than 0.01
GPa.  Our first-principles calculations show that the buckled phase of
the PbX monolayer is more stable than the centrosymmetric planar
phase~\cite{hanakata-PRB-96-161401-2017}, consistent with other DFT
studies~\cite{wan-AdvMat-29-1521-2017,
  kobayashi-surfaceScience-639-54-2015}. Detailed discussions on
the bistable nature, ferroelectric properties and orbital-spin texture
properties of lead chalcogenides can be found in our previous
paper~\cite{hanakata-PRB-96-161401-2017}. In the current work, the
deformations (atomic distortions) are applied to the optimized buckled
structure.

We used a finer grid for band structure calculations with the spin-orbit
interaction included.  We have tried several large numbers of $k$
points and found that a grid of 100 $k$ points between two symmetry
points (e.g between $X$ and $M$) is enough to obtain the Rashba
parameter $\lambda$ at the $M$ point~\cite{hanakata-PRB-96-161401-2017}. A
regular grid of 40$\times$40$\times$1 was used for the surface plot of
the energy spin splitting.

Here we tabulate the optimized (relaxed) geometrical parameters of
buckled PbX (X=S, Se, and Te) monolayers in
table~\ref{table:table1}. The Rashba parameters $\lambda$ are obtained
by taking the derivative of energy dispersion near the $M$ point. The
orbital coefficients are obtained by projecting the wave functions
into the atomic orbital basis. The unstrained values of $\lambda$ and
$\Delta$ are tabulated in table~\ref{table:table2}. From the table it
can be seen that the wave functions are mostly composed of in-plane
and out-of-plane of $p$ orbitals of Pb and an in-plane orbital of the
chalcogen X (X=S, Se, Te).
\begin{table}[h]
\small
  \caption{Relaxed lattice constant $a$, buckling angle $\theta$, buckling height $d_z$, nearest-neighbor bond distance $d$. }
\centering
\begin{tabular*}{0.5\textwidth}{@{\extracolsep{\fill}}lllll}
\hline 
& $a${\rm (\AA)} & $\theta(^\circ)$& $d_z${\rm (\AA)} & $d${\rm (\AA)} \\
\hline 
PbS   &   3.74 &   21.6 &  1.04  & 2.84 \\
PbSe    & 3.82 &  24.3    & 1.22    & 2.96 \\
PbTe    & 4.01   & 26.3     & 1.40 & 3.16 \\
\hline
\end{tabular*}
\label{table:table1}
\end{table}

\begin{table}[h]
\small
\caption{Rashba parameters $\lambda$, projected wave functions coefficients $|c_0|^2$, $|c_1|^2$, $|c_2|^2$ obtained from DFT
  and $\Delta$.  }
\centering
\begin{tabular*}{0.5\textwidth}{@{\extracolsep{\fill}}llllll}
\hline 
& $|c_0|^2$& $|c_1|^2$ & $|c_2|^2$ & $\lambda${\rm (eV\AA)} & $\Delta$ {\rm (eV)} \\
\hline 
PbS   &  0.305  & 0.534   & 0.115  & 3.40 & 5.36\\
PbSe  &  0.272 & 0.549  &  0.137  & 3.37 & 4.28\\
PbTe  &  0.286 & 0.522  &  0.130  & 3.18 & 3.83\\
\hline
\end{tabular*}
\label{table:table2}
\end{table}

\section{Electric polarization}
\label{sec:ferro}
We used the modern theory of
polarization~\cite{vanderbilt-PRB-47-1651-1993} to calculate the
spontaneous polarization implemented in the {\sc Quantum ESPRESSO}
package~\cite{QE-2009}. The electric polarization is calculated via
Berry phase calculation~\cite{vanderbilt-PRB-47-1651-1993}, which is
given by
\begin{equation} 
\vec{\mathcal{P}}=\frac{1}{V}\sum_{\tau}q^{\rm{ion}}_{\tau}{\bf R}_{\tau}-\frac{2i\rm{e}}{(2\pi)^3}\sum_{n}^{\rm{occ}}\int_{BZ}d^3{\bf k}e^{-i\vec{k}\cdot{\bf R}}\Big\langle \Psi_{n{\bf k}}\Big|\frac{\partial \Psi_{n{\bf k}}}{\partial {\bf k}}\Big\rangle, 
\label{eq:pol}
\end{equation}
where $q_\tau$ is the ionic charge plus the core electrons,
${\bf R}_{\tau}$ is the position of ions, $V$ is the unit cell
volume, $\rm{e}$ is the elementary charge, $n$ is the valence band
index, ${\bf k}$ is the wave vector, and $\Psi_{n{\bf k}}$ is the
electronic wave function. The first term is the contribution from core
electrons and ions, and the second term is the electronic contribution
defined as the adiabatic flow of current which can be calculated from
the Berry connection~\cite{vanderbilt-PRB-47-1651-1993}.

\begin{figure}
\includegraphics[width=8cm]{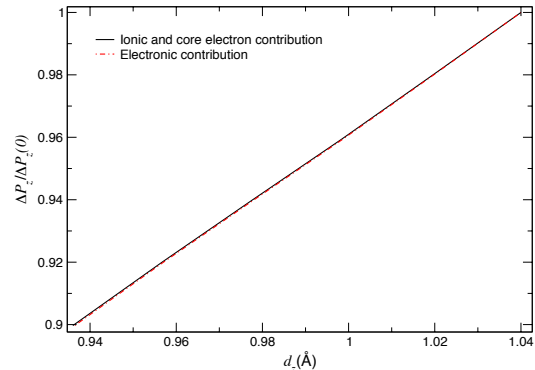}
\caption{From the DFT results we found that the ionic plus core
  electronic and the electronic (by Berry phase calculation)
  contributions are proportional to the distance between Pb and X
  (X=S, Se) in the $z$ direction.}
\label{fig:fig1S}
\end{figure}

The spontaneous polarization is calculated by taking the difference
between the polarization of the polar (buckled) state and the
non-polar (reference) state,
$\Delta \vec{\mathcal{P}}=\vec{\mathcal{P}}_{\rm
  polar}-\vec{\mathcal{P}}_{\rm non-polar}$.
To find the polarization at different heights, we change the
out-of-plane distance between the Pb and X (X=S, Se) atom while
keeping the in-plane lattice vectors fixed at the optimized buckled
values. It is a common practice to use a value on the order of the
bulk lattice constant (0.5--1 nm) to estimate the monolayer thickness
in order to compare the polarizations of monolayers to the typical
bulk ferroelectrics~\cite{hanakata-PRB-94-035304-2016,
  fei-PRL-117-097601-2016, wang-2DMat-4-015042-2017}. In this current
work we estimate the thickness to be 0.5 nm. In {\sc Quantum
  ESPRESSO}, spontaneous polarization with spin-orbit included can be
calculated using norm conserving pseudopotentials. A difference of
0.03 $\mu{\rm C/cm^{2}}$ is found when spin-orbit interaction is
included. Thus, to save computational time we only report spontaneous
polarization without inclusion of the spin-orbit interaction. This
small difference has also been reported
previously~\cite{stroppa-jpcl-6-2223-2015, leppert-jpcl-7-3683-2016}.
In figure.~\ref{fig:fig1S} we plot the polarization from the ionic
plus core electron contribution, and the electronic contribution, from
the Berry phase calculation, scaled by their values at zero strain as
a function of distance between the Pb and S atom in the $z$ direction.

\section{Tight binding}
\label{sec:tb}
The lead chalcogenide monolayer has two atoms per unit cell ($A, B$). Based
on density functional theories, the relevant orbitals near the valence
and conduction bands are $s$ and $p$ orbitals. The wave function of sublattice $A$ then can be written as 
\begin{equation}
\psi^{A}(r)=\frac{1}{\sqrt N}\sum_{{\bf k}, \mu}e^{i{\bf k}\cdot{\bf R}}a_{\mu, {\bf k}}\phi_\mu({\bf r}-{\bf R}), 
\end{equation}
where ${\bf R}$ is the lattice vector, ${\bf k}$ is a wave vector, $\mu$ is the
basis wave function
$[s, p_x, p_y, p_z]$.  Including
only nearest neighbor hopping the spinless Hamiltonian can be written
as
\begin{align}
H_{\rm orb}&=\sum_{\mu, \nu}\sum_{i, j}[t_{ij, \mu\nu}c^{\dagger}_{i, \mu}c_{j, \nu} + h.c] +\sum_{\mu, \nu}\sum_{i}E_{\mu\nu}c^{\dagger}_{i, \mu}c_{i, \nu},
\end{align}
where $\langle i, j\rangle$ runs over the onsite cell and the nearest
neighboring cells. $c^\dagger_{i, \mu}$ creates an electron in the
unit cell $i$ with atomic orbital $\mu$. We can write this more
compactly as
\begin{align}
H_{\rm orb }=\begin{pmatrix}H_{AA}& H_{AB}\\
H^\dagger_{AB} & H_{BB}
\end{pmatrix}\,,
\end{align}
where $H_{AA}$ (the onsite term) is given by 
\begin{align}
H_{AA}=\begin{pmatrix}E^s_A& 0 & 0 & 0\\
0& E^{p_x}_{A} & 0 & 0\\
0 & 0& E^{p_y}_{A} & 0\\
0 & 0 &  0& E^{p_z}_{A}\\
\end{pmatrix}\,.
\end{align} 
To write down the hopping matrix, we use the following Slater-Koster matrix elements for the orbitals of neighboring atoms~\cite{slater-PR-94-1498-1954}:
\begin{align}
&\text{$s$-$s$}: V_{ss\sigma}\,,
\nonumber
\\
&\text{$s$-$p$}: V_{sp\sigma}\hat{d}\cdot \hat{o}_j\,,
\nonumber
\\
&\text{$p$-$p$}: \left(\hat{o}_i\cdot\hat{o}_j\right) V_{pp\pi}+\left(\hat{o}_i\cdot\hat{d}\right)\left(\hat{o}_j\cdot\hat{d}\right)\left(V_{pp\sigma}-V_{pp\pi}\right)\,.
\label{eqn:SK_elements}
\end{align}
Here, $\hat{o}_i$ is the orientation of the $i$th orbital and $\hat{d}$ is the unit vector pointing from atom 1 to atom 2.
If we include up to first nearest neighbors only we can write the inter-lattice hopping matrix $H_{AB}\equiv K$ as
\begin{widetext}
\begin{align}
	K &= \Theta_\Gamma
	\begin{pmatrix}
		V_{ss\sigma}&0&0&-\gamma V_{sp\sigma}^{(1)}
		\\
		0&V_{pp\pi} + \alpha^2\Delta&0&0
		\\
		0&0&V_{pp\pi} + \alpha^2\Delta&0
		\\
		\gamma V_{sp\sigma}^{(2)}&0&0&V_{pp\pi} + \gamma^2\Delta
	\end{pmatrix}+
	4\alpha^2\Delta \Theta_M
	\begin{pmatrix}
		0&0&0&0
	\\
		0&0&1&0
	\\
		0&1&0&0
	\\
		0&0&0&0
	\end{pmatrix}
	\nonumber
	\\
        &+4\alpha \Theta_X
	\begin{pmatrix}
		0&i V_{sp\sigma}^{(1)}&0&0
	\\
		-i V_{sp\sigma}^{(2)}&0&0&-i\gamma\Delta
	\\
		0&0&0&0
	\\
		0&-i\gamma\Delta&0&0
	\end{pmatrix}
	+ 4\alpha \Theta_Y
	\begin{pmatrix}
		0&0&i V_{sp\sigma}^{(1)}&0
	\\
		0&0&0&0
	\\
		-i V_{sp\sigma}^{(2)}&0&0&-i\gamma\Delta
	\\
		0&0&-i\gamma\Delta&0
\end{pmatrix}\,.
\end{align}	
\label{eqn:K}
where
$\Theta_{\Gamma, M, X, Y}=\big[\cos\frac{k_xa}{2}\cos\frac{k_ya}{2},
\sin\frac{k_xa}{2}\sin\frac{k_ya}{2},
\sin\frac{k_xa}{2}\cos\frac{k_ya}{2},
\sin\frac{k_ya}{2}\cos\frac{k_xa}{2},
\big]$.
\end{widetext}
The momentum $\pi/a\leq k_{x/y}\leq\pi/a$ and $\gamma =
\sin\theta$.
To keep the expression more compact, we have introduced
$\Delta = V_{pp\sigma}-V_{pp\pi}$. In addition, since the $A$ and $B$
species are not necessarily the same, we have two quantities of the
$V_{sp\sigma}$ form.

While it is convenient to use $s$ and $p$ orbitals to write down the
hopping matrix, since we are interested in including SOI in our model,
it is helpful to go to a basis which is more natural for the angular
momentum operators:
\begin{align}
|0,0\rangle&=|s\rangle\,,
\quad
|1,\pm1\rangle=\frac{\mp|p_x\rangle-i|p_y\rangle}{\sqrt{2}}\,,
\quad
|1,0\rangle=|p_z\rangle\,,
\label{eqn:New_Basis}
\end{align}
where the first number represents the orbital momentum quantum number
and the second one is its projection along the $\hat z$
direction. This basis change does not alter the $H_{AA}$ and $H_{BB}$
matrices. The inter-lattice hopping portion of the Hamiltonian, on the
other hand, becomes
\begin{widetext}
\begin{align}
\bar K &=
\Theta_\Gamma \underbrace{\begin{pmatrix}V_{ss\sigma} &0&0&-\gamma V_{sp\sigma}^{(1)}
\\
0&V_{pp\pi}+\alpha^2\Delta&0&0
\\
0&0&V_{pp\pi}+\alpha^2\Delta&0
\\
\gamma V_{sp\sigma}^{(2)}&0&0&V_{pp\pi}+\gamma^2\Delta
\end{pmatrix}}_{K_{\Gamma}}+
4\alpha^2\Delta\Theta_M \underbrace{\begin{pmatrix}
0&0&0&0
\\
0&0&-i&0
\\
0&i&0&0
\\
0&0&0&0	
\end{pmatrix}}_{K_M}+
\nonumber
\\
&+2\sqrt{2}\alpha \Theta_X \underbrace{\begin{pmatrix}
0&-i V_{sp\sigma}^{(1)}&i V_{sp\sigma}^{(1)}&0
\\
i V_{sp\sigma}^{(2)}&0&0&i\gamma \Delta
\\
-i V_{sp\sigma}^{(2)}&0&0&-i\gamma\Delta
\\
0&i\gamma\Delta&-i\gamma\Delta&0
\end{pmatrix}}_{K_X}	
+2\sqrt{2}\alpha \Theta_Y \underbrace{\begin{pmatrix}
0& V_{sp\sigma}^{(1)}&V_{sp\sigma}^{(1)}&0
\\
V_{sp\sigma}^{(2)}&0&0&\gamma \Delta
\\
V_{sp\sigma}^{(2)}&0&0&\gamma\Delta
\\
0&-\gamma\Delta&-\gamma\Delta&0
\end{pmatrix}}_{K_Y}\,.
\label{eqn:Transformed_K}
\end{align}
\end{widetext}
From here we write $H\equiv U H_{\rm orb} U^{-1}$, where $U$ is a
matrix projector from the orbital basis to the angular momentum basis.

To include the SOI, we use the standard form describing the spin-orbit
coupling arising from the interaction with the nucleus:
\begin{equation}
H_\mathrm{SOI} = T_\mathcal{X}\left(\frac{L_+\otimes s_-+L_-\otimes s_+}{2}+L_z\otimes s_z\right)\,,
\label{eqn:SOI}
\end{equation}
where $\mathcal{X}$ is either Pb or X (X=S, Se, Te). The last term
modifies the diagonal elements of the self-energy for $|1,\pm1\rangle$
by adding (subtracting) $T_\mathrm{X}/2$ if $L_z$ and $s_z$ point in
the same (opposite) direction. The first tem couples
$|1,1\rangle\otimes |\downarrow\rangle$ with
$|1,0\rangle\otimes|\uparrow\rangle$ and
$|1,-1\rangle\otimes |\uparrow\rangle$ with
$|1,0\rangle\otimes|\downarrow\rangle$ with the coupling strength
$T_\mathrm{X}/\sqrt{2}$.

The total Hamiltonian can then be written as
\begin{equation}
H_{\rm tot}=H\otimes 1_{2x2}+H_{\rm SOI}
\end{equation}

\subsection{$M$ Point}
\label{sec:M_Point}
We first look around the $M$ point $k_x = k_y = \pi/a$.  To the leading order in
$q$, the hopping matrix $\tilde K$ is given by,
\begin{align}
\tilde K &=
4\alpha^2\Delta\begin{pmatrix}
0&0&0&0
\\
0&0&-i&0
\\
0&i&0&0
\\
0&0&0&0	
\end{pmatrix}
\nonumber
\\
&-a\sqrt{2}\alpha q\begin{pmatrix}
0&V_{sp\sigma}^{(1)}e^{-i\phi}& V_{sp\sigma}^{(1)}e^{i\phi}&0
\\
 V_{sp\sigma}^{(2)}e^{i\phi}&0&0&\gamma \Delta e^{i\phi}
\\
V_{sp\sigma}^{(2)}e^{-i\phi}&0&0&\gamma\Delta e^{-i\phi}
\\
0&-\gamma\Delta e^{-i\phi}&-\gamma\Delta e^{i\phi}&0
\end{pmatrix}	
\,,
\label{eqn:K_M}
\end{align}

where $\phi$ is the angle measured from the $\hat x$ direction. At $q=0$ ($k_x=k_y=\pi/a$), the Hamiltonian decomposes into several uncoupled blocks with the corresponding bases:
\begin{align}
	H_s^{m,\pm} &= E^s_m: |0,0\rangle\otimes|\pm\rangle\otimes|m\rangle\,,
	\nonumber
	\\
	H_p^{mn,\pm}& =\begin{pmatrix}
		E^p_m-\frac{T_m}{2}&\frac{T_m}{\sqrt{2}}&\mp 4i\alpha^2\Delta
		\\
		\frac{T_m}{\sqrt{2}}&E^p_m&0
		\\
		\pm 4i\alpha^2\Delta&0&E^p_n+\frac{T_n}{2}
		\end{pmatrix}:
		\begin{pmatrix}
			|m\rangle\otimes|1,\pm1\rangle\otimes|\mp\rangle
			\\
			|m\rangle\otimes|1,0\rangle\otimes|\pm\rangle
			\\
			|n\rangle\otimes|1,\mp1\rangle\otimes|\mp\rangle
		\end{pmatrix}\,,
\label{eqn:H_M}
\end{align}
where $m\neq n$ labels the sublattices and the middle $|\pm\rangle$ ket
denotes the spin state. Using the direct sum notation, we can write
down the total Hamiltonian as
$H = H_s^{A,+}\oplus H_s^{A,-}\oplus H_s^{B,+}\oplus H_s^{B,-}\oplus
H_p^{AB,+}\oplus H_p^{AB,-} \oplus H_p^{BA,+}\oplus H_p^{BA,-}$.

From $H_s$, we see that for a given $m$, the eigenstates are
spin-degenerate. The degeneracy becomes four-fold if the atoms of
sublattices $A$ and $B$ are the same, leading to
$E^p_A = E^p_B$. Equation~\eqref{eqn:K_M} shows that
at finite $q$ there is no coupling between the degenerate
$|0,0\rangle$ states that is linear in momentum. This means that the
bands composed of $s$ orbitals have local extrema at the $M$ point.

Next, we turn to $H_p$ from Eq.~\eqref{eqn:H_M}. Just like for $H_s$,
the bands are doubly or four-fold degenerate depending on whether the
sublattices are composed of the same atomic species. Without making
assumptions about the lattice composition, the general form of the
degenerate states is
\begin{align}
	|\Psi^\pm\rangle_{mn} &=c_0|m\rangle\otimes|1,\pm1\rangle\otimes|\mp\rangle+c_1|m\rangle\otimes|1,0\rangle\otimes|\pm\rangle
	\nonumber
	\\
	&\pm ic_2 |n\rangle\otimes|1,\mp1\rangle\otimes|\mp\rangle\,,
	\label{eqn:Psi_M}
\end{align}
with $c_0$, $c_1$, and $c_2$ real. At finite $q$,
\begin{equation}
	\,_{mn}\langle \Psi^+| H|\Psi^-\rangle_{mn} =-a\sin2\theta  c_1 c_2\left(\Delta iqe^{-i\phi}\right)\varepsilon_{mn}\,,
	\label{eqn:PsiPsi_M}
\end{equation}
where $\varepsilon_\mathrm{AB} = -\varepsilon_\mathrm{BA} = 1$ is the
two-dimensional Levi-Civita symbol. This coupling between the
degenerate states leads to an effective Rashba-like Hamiltonian:
\begin{equation}
	H_\mathrm{eff}^{mn} = a\sin2\theta c_1c_2\Delta\varepsilon_{mn}\left[\left(\mathbf{q}\times\pmb{\sigma}\right)\cdot\hat z\right]:\begin{pmatrix}
		|\Psi^+\rangle_{mn}
		\\
		|\Psi^-\rangle_{mn}
	\end{pmatrix}\,.
	\label{eqn:H_eff_M}, 
\end{equation}
or in the matrix form 
\begin{align}
H_\mathrm{eff} &=\begin{pmatrix}
0&-i\lambda(q_x-iq_y)\\
i\lambda(q_x+iq_y)&0
\end{pmatrix}\,,
\label{eqn:Heff}
\end{align}

We use values of $c_0$, $c_1$, and $c_2$ obtained from DFT results. To
give better physical pictures of these coefficients, we will solve
the Hamiltonian Eq.~\ref{eqn:H_M}. We treat the spin orbit
interaction (SOI) as perturbations and we will assume that $T_m\gg T_n$ where
$m$ is the index denoting Pb with strong SOI and $n$ denotes weak SOI
of chalcogen atom.  Focusing on $H_p^{mn,+}$, Eq.~\ref{eqn:H_M} becomes 
\begin{equation}
	H_p^{mn,+}=\begin{pmatrix}
		E^p_m&0&-4i\alpha^2\Delta
		\\
		0&E^p_m&0
		\\
		4i\alpha^2\Delta&0&E^p_n
		\end{pmatrix}:
		\begin{pmatrix}
			|m\rangle\otimes|1,1\rangle\otimes|-\rangle
			\\
			|m\rangle\otimes|1,0\rangle\otimes|+\rangle
			\\
			|n\rangle\otimes|1,-1\rangle\otimes|-\rangle
		\end{pmatrix}\,,
\label{eq:H_un}
\end{equation}
and the perturbation 
\begin{equation}
	\delta H_p^{mn,+}=\begin{pmatrix}
	-\frac{T_m}{2}&\frac{T_m}{\sqrt{2}}&0
        \\
        \frac{T_m}{\sqrt{2}}&0&0
          \\
          0&0&0
		\end{pmatrix}:
		\begin{pmatrix}
                  	|m\rangle\otimes|1,1\rangle\otimes|-\rangle
			\\
			|m\rangle\otimes|1,0\rangle\otimes|+\rangle
			\\
			|n\rangle\otimes|1,-1\rangle\otimes|-\rangle
		\end{pmatrix}\,.
\label{eq:H_per}
\end{equation}
We first solved Eq.~\ref{eq:H_un} to find the eigenvalues and
eigenvectors and used first order perturbation theory to obtain the
corrections to the eigenvectors. Using MATHEMATICA, we found to the first order
in $T_m$ that
\begin{equation}
|c_1c_2|\simeq\frac{T_m(E^p_m-E^p_n+\sqrt{(E^p_m-E^p_n)^2+64\alpha^4\Delta^2})}{8\sqrt{2}\alpha^2\Delta(E^p_n-E^p_m+\sqrt{(E^p_m-E^p_n)^2+64\alpha^4\Delta^2})}.
\label{eq:c1c2}
\end{equation}
Recall that we defined Rashba parameter
$\lambda\equiv a \sin2\theta \Delta c_1c_2$. From Eq.~\ref{eq:c1c2} we
see that $|c_1c_2|$ weakly depends on strains. For this reason, in the
main text we assumed $c_1$ and $c_2$ are constant and the corrections
to $\lambda$ come mostly from $\Delta$ and $\theta$.

\section{Magnetic Field}
\widetext
\label{sec:MagneticField}

Let us now try to include external fields to the system. The magnetic
field can be included via the Peierls substitution so that
$\mathbf{q}\rightarrow \mathbf{q} - e\mathbf{A}/c$, where $\mathbf{A}$
is the vector potential. In addition, applying an external magnetic
field leads to the interaction of the electron angular momentum with
the field.

The total magnetic moment of an electron is given by
\begin{equation}
	\mathbf{\mu} = -\mu_B\frac{\mathbf{L}+2\mathbf{S}}{\hbar}\,,
\end{equation}
so that
\begin{equation}
	\mathbf{B}\cdot\mu =-\mu_B\frac{B_x\left(\frac{L_++L_-}{2}+S_++S_-\right)+B_y\left(\frac{L_+-L_-}{2i}+\frac{S_+-S_-}{i}\right)+B_z\left(L_z+2S_z\right)}{\hbar}\,.
\end{equation}
Setting $\mathbf{B} = \left(B_\parallel\cos\tau,B_\parallel\sin\tau,B_\perp\right)$ gives
\begin{align}
	\mathbf{B}\cdot\mu &=-\mu_B\frac{B_\parallel\cos\tau\left(\frac{L_++L_-}{2}+S_++S_-\right)-iB_\parallel\sin\tau\left(\frac{L_+-L_-}{2}+S_+-S_-\right)+B_\perp\left(L_z+2S_z\right)}{\hbar}
	\nonumber
	\\
	&=-\mu_B\frac{B_\parallel\left[e^{-i\tau}\left(\frac{L_+}{2}+S_+\right)+e^{i\tau}\left(\frac{L_-}{2}+S_-\right)\right]+B_\perp\left(L_z+2S_z\right)}{\hbar}\,.
\end{align}
The first term $\propto B_\parallel$ introduces coupling between $|\Psi^\mathrm{+/-}\rangle$ while the last term $\propto B_\perp$ modifies and breaks the symmetry between the degenerate states. Starting with last term, we get
\begin{equation}
	\langle \Psi^+|\mathbf{B}\cdot\mu|\Psi^+\rangle = -\langle \Psi^-|\mathbf{B}\cdot\mu|\Psi^-\rangle = -\mu_BB_\perp\left(c_1^2-2c_2^2\right)\,.
\end{equation}

Next, we apply the first term onto $|\Psi^+\rangle$:
\begin{align}
	&-\mu_B\frac{B_\parallel\left[e^{-i\tau}\left(\frac{L_+}{2}+S_+\right)+e^{i\tau}\left(\frac{L_-}{2}+S_-\right)\right]}{\hbar}(
c_0|m\rangle\otimes|1,1\rangle\otimes|-\rangle+c_1|m\rangle\otimes|1,0\rangle\otimes|+\rangle+ic_2|n\rangle\otimes|1,-1\rangle\otimes|-\rangle)=
	\nonumber
	\\
	=&-\mu_B B_\parallel\Big[e^{-i\tau}c_0|m\rangle|1, 1\rangle|+\rangle+e^{-i\tau}\frac{c_0}{\sqrt{2}}|m\rangle|1, 0\rangle|-\rangle +e^{-i\tau}\frac{c_1}{\sqrt{2}}|m\rangle|1, 1\rangle|+\rangle
	\nonumber
	\\
	&+e^{i\tau}\frac{c_1}{\sqrt{2}}|m\rangle|1, -1\rangle|+\rangle+e^{i\tau}c_1|m\rangle|1, 0\rangle|-\rangle+e^{-i\tau}i\frac{c_2}{\sqrt{2}}|n\rangle|1, 0\rangle|-\rangle+e^{i\tau}ic_2|n\rangle|1, -1\rangle|+\rangle\Big]
	\nonumber
	\\
	=&-\mu_B B_\parallel\Big[e^{i\tau}(\frac{c_0}{\sqrt{2}}+c_1)|m\rangle|1, 0\rangle|-\rangle+e^{-i\tau}(c_0+\frac{c_1}{\sqrt{2}})|m\rangle|1, 1\rangle|+\rangle +c_1e^{i\tau}c_1|m\rangle|1, -1\rangle|+\rangle
	\nonumber
         \\
        &+e^{-i\tau}i\frac{c_2}{\sqrt{2}}|n\rangle|1, 0\rangle|-\rangle+e^{i\tau}ic_2|m\rangle|1, -1\rangle|+\rangle\Big]
	\label{eqn:B_parallel_Coupling}
\end{align}
Now, we apply $\langle\Psi^-|$ onto Eq.~\eqref{eqn:B_parallel_Coupling}. We see that the states on $|n\rangle$ drop out. The remaining states yield
\begin{equation}
	\langle\Psi^-|\mathbf{B}\cdot\mu|\Psi^+\rangle = -\mu_B B_\parallel e^{i\tau}\left[c_1\left(\frac{c_0}{\sqrt{2}}+c_1+c_0\right)\right]\,.
\end{equation}

Thus, our general Hamiltonian becomes
\begin{equation}
	H =\lambda\left[\left(\mathbf{q}-\frac{e\mathbf{A}}{c}\right)\times\pmb{\sigma}\right]\cdot \hat{z}+m_\perp\sigma_zB_\perp+m_\parallel\mathbf{B}_\parallel\cdot\sigma_\parallel\,
\end{equation}
where $m_\perp=-\mu_B(c_1^2-2c_2^2)$ and $m_\parallel=-\mu_B\left [c_1\left(\frac{c_0}{\sqrt{2}}+c_1+c_0\right)\right]$

\section{In-Plane Field}
\label{sec:InPlaneBField}
If the field is in-plane, the Hamiltonian is given by
\begin{equation}
	H = \begin{pmatrix}
 			0 & i\lambda qe^{i\phi}+\mathcal{B} e^{-i\tau}
 			\\
 			-i\lambda qe^{-i\phi}+\mathcal{B} e^{i\tau}&0
 		\end{pmatrix} = \lambda\begin{pmatrix}
 			0 & iqe^{i\phi}+\frac{\mathcal{B}}{\lambda} e^{-i\tau}
 			\\
 			-iqe^{-i\phi}+\frac{\mathcal{B}}{\lambda} e^{i\tau}&0
 		\end{pmatrix}\,,
\end{equation}
where we have defined $\mathcal{B}\equiv m_{\parallel}B_\parallel$.
The eigenvalues become
\begin{align}
	\mathcal{E} &=\pm \lambda \sqrt{\left(q_x-\frac{\mathcal{B}}{\lambda}\sin\tau\right)^2+\left(q_y-\frac{\mathcal{B}}{\lambda}\cos\tau\right)^2} \,.
\end{align}
Applying an in-plane magnetic field shifts the cone in the Brillouin zone.

Let us take a closer look at the Hamiltonian:
\begin{align}
	H &= \lambda\begin{pmatrix}
 			0 & i\left[\left(q_x-\frac{\mathcal{B}}{\lambda}\sin\tau\right)+i\left(q_y-\frac{\mathcal{B}}{\lambda}\cos\tau\right)\right]
 			\\
 			-i\left[\left(q_x-\frac{\mathcal{B}}{\lambda}\sin\tau\right)-i\left(q_y-\frac{\mathcal{B}}{\lambda}\cos\tau\right)\right]&0
 		\end{pmatrix}=
 		\nonumber
 		\\
 		&= \lambda\begin{pmatrix}
 			0 & ipe^{i\xi}
 			\\
 			-ipe^{-i\xi}&0
 		\end{pmatrix}\,.
\end{align}
The eigenstates are
\begin{align}
	|I\rangle &= \frac{|\Psi^\mathrm{+}\rangle+ie^{-i\xi}|\Psi^\mathrm{-}\rangle}{\sqrt{2}}\,,
	\nonumber
	\\
	|II\rangle &= \frac{|\Psi^\mathrm{+}\rangle-ie^{-i\xi}|\Psi^\mathrm{-}\rangle}{\sqrt{2}}\,.
\end{align}

Now we can obtain the in-plane spin texture for the cones. First, it is easy to show that
\begin{equation}
	\langle \Psi^\mathrm{I}|\sigma_{x/y}|\Psi^\mathrm{I}\rangle = \langle \Psi^\mathrm{II}|\sigma_{x/y}|\Psi^\mathrm{II}\rangle = 0\,.
\end{equation}
Next,
\begin{align}
	\langle \Psi^\mathrm{II}|\sigma_{x}|\Psi^\mathrm{I}\rangle &= c_1^2 \langle + |\sigma_x|-\rangle = c_1^2\,,
	\nonumber
	\\
	\langle \Psi^\mathrm{II}|\sigma_{y}|\Psi^\mathrm{I}\rangle &= c_1^2 \langle + |\sigma_y|-\rangle\ = -ic_1^2\,.
\end{align}
This leads to
\begin{align}
	\langle I|\sigma_x|I\rangle &= \frac{-ie^{i\xi}}{2}c_1^2+\frac{ie^{-i\xi}}{2}c_1^2=-i\frac{c_1^2}{2}\left(e^{i\xi}-e^{-i\xi}\right) = c_1^2\sin\xi\,,
	\nonumber
	\\
	\langle I|\sigma_y|I\rangle &=-ic_1^2\frac{-ie^{i\xi}}{2}+ic_1^2\frac{ie^{-i\xi}}{2} = -c_1^2\frac{e^{i\xi}}{2}-c_1^2\frac{e^{-i\xi}}{2} = -c_1^2\cos\xi\,,
	\nonumber
	\\
	\langle II|\sigma_x|II\rangle &=\frac{ie^{i\xi}}{2}c_1^2+\frac{-ie^{-i\xi}}{2}c_1^2 = -c_1^2\sin\xi\,,
	\nonumber
	\\
	\langle II|\sigma_y|II\rangle &=ic_1^2\frac{-ie^{i\xi}}{2}-ic_1^2\frac{ie^{-i\xi}}{2} =c_1^2\cos\xi\,.
\end{align}
As a result, the spin texture becomes:
\begin{align}
	\langle I|\hat\sigma|I\rangle &= c_1^2\left(\hat{x}\sin\xi-\hat{y}\cos\xi\right)\propto \left(\hat{x}p_y-\hat{y}p_x\right)\,,
	\nonumber
	\\
	\langle II|\hat\sigma|II\rangle &= -c_1^2\left(\hat{x}\sin\xi-\hat{y}\cos\xi\right)\propto -\left(\hat{x}p_y-\hat{y}p_x\right)\,.
\end{align}
Recall that 
\begin{align}
	p_x = q_x-\frac{\lambda}{\hbar v}\sin\tau\,,
	\nonumber
	\\
	p_y = q_y-\frac{\lambda}{\hbar v}\cos\tau\,.
\end{align}
This means that spin contours now revolve not around the
$\mathbf{q}=0$ point but instead around a $\mathbf{p}=0$ point.

\end{document}